\documentclass[aps,pra,showpacs,eqsecnum,twocolumn]{revtex4}
\usepackage{graphicx}
\usepackage{amsmath,amssymb,amsthm,dsfont,bm}

\usepackage{color}

\begin{document}
\preprint{}
\title{Breaking the weak Heisenberg limit}
\author{Alfredo Luis}
\email{alluis@fis.ucm.es}
\homepage{http://www.ucm.es/info/gioq}
\affiliation{Departamento de \'{O}ptica, Facultad de Ciencias
F\'{\i}sicas, Universidad Complutense, 28040 Madrid, Spain }
\date{\today}

\begin{abstract}
We  provide a very simple case showing that the weak form of the Heisenberg limit can be beaten 
while the prior information is improved without bias. 
\end{abstract}

\maketitle

\section{Introduction}

The question of the ultimate quantum limits to the precision in signal detection has received a 
great deal of  attention. One of the reasons for this interest relies on the growing importance of 
novel quantum technologies where quantum physics is applied to an increasing number of 
practical tasks. 

In this work we focus on the most common scheme in quantum metrology, where the signal 
to be detected is encoded as a shift $\phi$ of the phase of an harmonic oscillator. In the most 
practical terms this means an electromagnetic field mode illuminating an interferometer.

A key issue in quantum metrology is the trade-off between resolution and resources employed, 
usually counted as the number of photons in the probe state.
The common belief points to an ultimate minimum uncertainty $\Delta \phi $ scaling as 
the inverse of the total number of photons $N$, this is $\Delta \phi_s \propto  1/N$
\cite{GLM04,GLM11,MT12,GLM12,BHZW12,HBZW12,RN12,GM12,GLM12b,HW12,GL12}. 
We will refer to this as the {\it strong form of the Heisenberg limit} or simply {\it strong 
Heisenberg limit}. In order to reach this limit all the photons must be employed in a single 
realization of the measurement and a highly nonclassical probe state with a very large 
number of photons is required. Such states are very difficult to generate and extremely fragile 
against practical imperfections \cite{HMPEPC97,DBS13,DJK15}. These  considerations 
may spoil the actual practical meaning of the strong Heisenberg limit. 

In a more practical scenario we should consider instead the repetition $m$ times of the 
measurement with identical probes prepared in a  nonclassical state with small mean 
number $\bar{n}$, such that $N =  m \bar{n}$ can be still very large. Considering that 
the $m$ repetitions are statistically independent, the minimum uncertainty would scale 
as  \cite{PHS15}
\begin{equation}
\textrm{Weak form:} \quad  \Delta \phi_w \propto \frac{1}{\sqrt{m} \bar{n} },
\end{equation}
which is rather different from the strong form 
\begin{equation}
\textrm{Strong form:}  \quad \Delta \phi_s \propto \frac{1}{m \bar{n}} = \frac{1}{N} ,
\end{equation}
specially for meaningful situations where $m$ will be far larger than $\bar{n}$. We will refer to 
$\Delta \phi_w$ as the {\it weak form of the Heisenberg limit}, or simply  {\it weak 
Heisenberg limit}.

The actual meaning of the strong Heisenberg limit has been much debated
\cite{SSW89,SS91,ZS92,ZH92,BLC92,SB92,LBC93,SB94,ARCPHLD10,RL12,LR13,AL13,ZJCLF13,DKG12,LP13}.
However, the weak Heisenberg limit has not so extensively examined \cite{SB94,BHZW12,LP13}, 
although it has a more deep practical meaning as discussed above. 

In this work we show by means of an extremely simple example the meaningful beating of the weak
form of  the Heisenberg limit. Moreover,  this example suggests that the strong limit may be as well 
approached in this same scenario of large $m$ and small $\bar{n}$. 

We can  benefit from many conclusions of the strong-limit scenario. To begin with, one 
must be careful concerning the performance estimators we can trust \cite{SSW89,SS91,ZS92,ZH92,BLC92,SB94}. 
It is crucial to check the presence of bias and whether meaningful improvement over the prior information 
is achieved \cite{RL12,MT12,GM12,HW12,LP13}. Because of this we mainly focus on Bayesian-like approaches 
involving averages over the prior knowledge. Nevertheless,  we will contrast the results also with  point-wise 
approaches such as the  Cram\'{e}r-Rao bounds. 
 
\section{Detection scheme}
The physical system for the detection is a single-mode electromagnetic field. The probe is 
prepared in the state $| \psi \rangle$ expressed in the photon-number basis as 
\begin{equation}
\label{probe}
| \psi \rangle = \sqrt{1 -\nu^2} |0 \rangle + \nu | \bar{n}/\nu^2 \rangle ,
\end{equation}
where $\nu$ is a parameter that will be considered small enough $\nu \ll 1$, while $\bar{n}$
is the mean number of photons of each probe-state realization, assuming always that  
$\bar{n}/\nu^2$ is an integer. This kind of states have been considered before \cite{SB94,BHZW12} 
and known for example as unbalanced cat states \cite{KPNDM14}.

In this scheme the  signal to be detected induces a phase shift $\phi$ transforming the probe state into 
\begin{equation}
\label{ps}
| \psi (\phi ) \rangle = \sqrt{1 -\nu^2} |0 \rangle + e ^{i \phi \bar{n} /\nu^2} \nu | \bar{n}/\nu^2 \rangle .
\end{equation}
We will consider that all what is known about the signal is that  $\phi$ is comprised in the interval $[0,W]$.  
This  knowledge is often summarized in a {\it prior distribution} $P ( \phi )$, in our case as $P ( \phi ) = 1/W$ 
for $\phi \in [0,W]$ and $P ( \phi ) = 0$ otherwise.

\section{Bounds}

In general terms many variable factors affect the estimation performance, such as probe state, signal 
codification, measurement performed, data analysis followed, experimental imperfections, and so on. 
Because of this, most performance analysis focus on the derivation of lower bounds on the estimation 
error, rather than dealing with exact precision limits. This is clearly discussed in Ref. \cite{JD15}, where 
several lower bounds to the estimation uncertainty $\Delta \tilde{\phi}$ are presented and discussed. 

Among them, the most popular is the Cram\'{e}r-Rao bound, that focus on the variance of the signal 
estimator $\tilde{\phi}$
\begin{equation}
\Delta^2 \tilde{\phi} =  \sum_k P ( k | \phi )  \left [ \tilde{\phi} (k) - \phi \right ]^2 ,
\end{equation} 
 where $k$ are the outcomes of the measurement performed, assumed discrete for simplicity, 
$P ( k | \phi )$ the probability of outcome $k$ when the signal is $\phi$, and $\tilde{\phi} (k) $ 
is the estimation of $\phi$ after the outcome $k$. Then it can be seen that
\begin{equation}
\label{CRB}
\Delta^2 \tilde{\phi} \geq \Delta^2 \tilde{\phi}_{CR} = \frac{1}{m F}, 
\end{equation}
where $F$ is the Fisher information of a single measurement,
\begin{equation}
\label{FI}
F = \sum_k \frac{1}{P ( k | \phi )} \left ( \frac{\partial 
P ( k | \phi )}{\partial \phi} \right )^2 .
\end{equation}

The dependence of $\Delta \tilde{\phi}_{CR}$ on the measurement performed can be removed leading 
to the quantum Cram\'{e}r-Rao bound $ \Delta \tilde{\phi}_{QCR}$,  which is again of the form (\ref{CRB}) 
but where the Fisher information (\ref{FI}) is replaced by the quantum Fisher information $F_Q$. For probes 
in pure states $F_Q$ reads simply as the variance of the generator $G$ of the phase 
shift  on the probe state
\begin{equation}
\label{QFI}
F_Q =  4 \left ( \langle \psi (\phi ) | G^2 | \psi (\phi ) \rangle -
\langle \psi (\phi ) | G | \psi (\phi ) \rangle^2 \right ) ,
\end{equation}
and in our case $G$ is the number operator $G = \hat{n}$. Note that in general these point-wise bounds 
depend on the unknown signal value $\phi$, ignore prior information $P(\phi)$, and do not address 
whether the estimation protocol is efficient reaching the lower bound.

An alternative picture that can take into account all these points is provided by  Bayesian approaches that 
aim constructing a posterior distribution for the signal estimator $\tilde{\phi} $, for example in the form
\begin{equation}
\label{post}
P  ( \tilde{\phi} | \phi ) \propto P \left [ k  ( \tilde{\phi} ) | \phi \right ] ,
\end{equation} 
where $k  ( \tilde{\phi} )$ is given by inverting the relation $\tilde{\phi} (k)$. A clear advantage of this 
approach is that solves many issues at once, such as unbiasedness and efficiency. Within this Bayesian 
scenario the uncertainty on $\tilde{\phi}$ can be estimated, including the prior information contained 
in $P(\phi)$, for example  via the mean square estimation error averaged over the prior distribution $P ( \phi)$
\begin{equation}
\Delta^2 \tilde{\phi} = \int d\phi \sum_k P ( k | \phi ) P ( \phi) \left [ \tilde{\phi} (k) - \phi \right ]^2 .
\end{equation}
A suitable lower bound for $\Delta^2 \tilde{\phi}$ is  Ziv-Zakai bound \cite{MT12}
\begin{equation}
\label{ZZ}
\Delta^2 \tilde{\phi} \geq \frac{1}{2} \int_0^W d\phi \phi \left ( 1- \frac{\phi}{W} \right ) \left [ 
1- \sqrt{1 -\left | \langle \psi | \psi (\phi) \rangle \right |^{2m}} \right ]  ,
\end{equation}
where 
\begin{equation}
\left | \langle \psi | \psi (\phi) \rangle \right |^2 = \left ( 1 - \nu^2 \right )^2 + \nu^4 +
2 \nu^2 \left ( 1 - \nu^2 \right ) \cos \left ( \bar{n} \phi/\nu^2 \right ).
\end{equation}
In order to deal with practicable expressions let us assume, as a first condition, that 
\begin{equation}
\label{1c}
 W \bar{n} /\nu^2 \ll 1 \quad \textrm{1st condition} ,
\end{equation}
so that in due course we may approximate $\cos \left ( \bar{n} \phi/\nu^2 \right )$ as $\cos x \simeq 
1-x^2/2$ as well as   $(1- y)^m \simeq e^{- my}$ within the range of values allowed for $\phi$ by the 
prior distribution. In such a case 
\begin{equation}
\label{ap}
\left | \langle \psi | \psi (\phi) \rangle \right |^{2m} \simeq \left ( 1 -  \bar{n}^2 \phi^2 /\nu^2 \right )^m 
\simeq e^{-  m \bar{n}^2 \phi^2 /\nu^2}.
\end{equation}
Moreover, in order to proceed with user-friendly expressions we further assume as a second 
condition that 
\begin{equation}
\label{2c}
\sqrt{m}\bar{n} W /\nu \gg 1 \quad \textrm{2nd condition} .
\end{equation}
This is natural since we expect that a very large number of repetitions $m$ is needed to reach a 
meaningful resolution. Thus we may approximate  in Eq. (\ref{ZZ}) $1-\sqrt{1-z} \simeq z/2$ with 
$z =  \exp (-  m \bar{n}^2 \phi^2 /\nu^2)$. Finally, after the $\phi$ integration we have 
\begin{equation}
\label{bwHl}
\Delta^2 \tilde{\phi} \geq \Delta^2 \tilde{\phi} _{ZZ} = \frac{\nu^2}{8 m \bar{n}^2}  = \frac{m \nu^2}{8 N^2}  .
\end{equation}
This is the final form for the Ziv-Zakai bound. The first conclusion is that this scheme 
clearly includes the possibility of beating the weak Hesienberg limit thanks to the 
small factor $\nu$ in the numerator of $\Delta \tilde{\phi}_{ZZ} $. Moreover, now we get that the 
condition (\ref{2c}) means that the bound is clearly smaller than the prior information, 
$ \Delta \tilde{\phi}_{ZZ} \ll W $. 

The conjunction of the two above conditions (\ref{1c}) and (\ref{2c}) implies that $m \nu^2 \gg 1$, 
which agrees with the results in Ref.  \cite{MT12} regarding the strong Heisenberg limit, since this 
would imply that $\Delta \tilde{\phi}_{ZZ} \gg \Delta \tilde{\phi}_s $. Nevertheless, we recall that 
conditions (\ref{1c}) and (\ref{2c}) were imposed in order to get manageable expressions. But 
the scheme  can work equally  well without them, as demonstrated by  the following example. 

After Eq. (\ref{bwHl}) it is not excluded that this strategy may approach the strong Heisenberg limit $\Delta 
\tilde{\phi}_{ZZ} \simeq \Delta \tilde{\phi}_s $ provided that $m \nu^2 \simeq 1$.
This suggests that the attainability of the strong limit scaling refers essentially to the $\bar{n}$ scaling.
The other variables $m, \nu$ are customarily fixed as function of  $\bar{n}$ to obtain the desired result 
for $\Delta \tilde{\phi}$ up to constant factors.

Finally, as shown in Ref. \cite{JD15} there is Bayesian Cram\'{e}r-Rao bound suitably mixing both 
strategies in the form
\begin{equation}
\label{BCR}
\Delta^2 \tilde{\phi}_{BCR} = \frac{1}{m \int d \phi P(\phi)  F + \mathcal{I} }, 
\end{equation}
where $F$ is the Fisher information associated to the measurement, while $\mathcal{I}$ represents Fisher 
information corresponding to the prior information
\begin{equation}
\mathcal{I}  = \int d \phi  \frac{1}{P ( \phi )} \left ( \frac{\partial 
P ( \phi )}{\partial \phi} \right )^2 .
\end{equation} 

\section{Posterior distribution}

The above result (\ref{bwHl}) for the Ziv-Zakai bound does not prove that the weak Heisenberg limit 
can be actually beaten in a specific scheme since this is a bound that may or may not be reached. 
In this section we proceed by showing an specific detection scheme providing a proof of principle 
that the bound  can be actually approached. We also compare the resolution reached with the 
other bounds in Sec. III as a consistence check. To this end we begin with by deriving explicitly the 
{\it posterior distribution} $P  ( \tilde{\phi} | \phi )$ for the estimated phase $\tilde{\phi}$ conditioned 
to the true unknown phase shift $\phi$ as presented in Eq. (\ref{post}).

The measurement in this example has just two outputs labelled $\pm$ whose probabilities 
$P (\pm | \phi) $ are given by projecting the signal-transformed state $| \psi (\phi ) \rangle$ in 
Eq. (\ref{ps}) on the orthogonal vectors $| \pm \rangle $, expressed in the number basis as
\begin{equation}
| \pm \rangle = \frac{1}{\sqrt{2}} \left ( | 0 \rangle \pm i |\bar{n}/\nu^2 \rangle \right ) .
\end{equation}
This is 
\begin{equation}
\label{ppme}
P ( \pm | \phi ) = \frac{1}{2} \pm  \nu \sqrt{1-\nu^2} \sin \left ( \phi \bar{n}/\nu^2 \right ) .
\end{equation}
When repeating the measurement $m$ times, the probability that we get $k$ positive 
outcomes and $m-k$ negatives is the binomial
\begin{equation}
\label{bin}
P  ( k | \phi  ) = \begin{pmatrix} m \\ k  \end{pmatrix}  P^k (+| \phi) P^{m-k} (- | \phi) .
\end{equation}
In the limit of large $m$, which is our case here, the binomial  (\ref{bin}) can be well approximated 
by the  Gaussian
\begin{equation}
\label{bG}
P  ( k | \phi  ) \simeq \frac{ \exp \left [ - \frac{ (k-m P (+ | \phi) )^2}{2 m P (+ | \phi) P (- | \phi)} \right ] }
{\sqrt{2 \pi m P (+ | \phi) P (- | \phi)}} .
\end{equation}
Following a maximum likelihood strategy, to each result with $k$ positive 
outcomes we can assign the estimator $\tilde{\phi}$ given by relation 
\begin{equation}
\label{est}
P (+ | \tilde{\phi} ) = k/m ,
\end{equation}
as the phase shift that maximizes the probability $P  ( k | \phi  )$ of obtaining the result actually obtained. 
Using again that the first condition (\ref{1c}) holds
\begin{equation}
\label{ppma}
P (+ | \tilde{\phi} ) \simeq \frac{1}{2} \pm  \phi \bar{n}/\nu  = k/m,
\end{equation}
so that 
\begin{equation}
\label{di}
\tilde{\phi} = \frac{k-m/2}{m \bar{n} /\nu} , \quad k (\tilde{\phi} ) = m/2+m \bar{n} \tilde{\phi} /\nu .
\end{equation}
With this, the final posterior distribution becomes after Eqs.  (\ref{post}), (\ref{bG}) and  (\ref{di}) is
\begin{equation}
\label{posta}
P  ( \tilde{\phi} | \phi ) \simeq \sqrt{\frac{2 m \bar{n}^2}{\pi \nu^2}}
\exp \left [ -2 \frac{ m \bar{n}^2}{\nu^2} \left ( \tilde{\phi} - \phi \right )^2 \right ]  ,
\end{equation} 
where we have approximated $P (+ | \phi) P (- | \phi) \simeq 1/4$. Note that $P ( \tilde{\phi} | \phi )$ 
readily provides the uncertainty of the estimator as
\begin{equation}
\label{ue}
\Delta^2 \tilde{\phi}  \simeq \frac{\nu^2}{4 m \bar{n}^2}  ,
\end{equation}
which is just twice the Ziv-Zakai bound (\ref{bwHl}). We can check also in Eq. (\ref{posta}) that in this limit 
there is no bias, since the mean value of the estimator $\tilde{\phi}$ coincides with the true value $\phi$.

Therefore this scheme is able to beat the weak Heisenberg limit for a suitable choice of large $m$ and
small $\nu$. At difference with the strong Heisenberg form,  in this case the violation is not jeopardized 
by any relation between $m$ and $\nu$. The only requirement  is that $m \gg 1$ and $\nu \ll 1$. In the next 
section we present a specific numerical example.

Regarding the Cram\'{e}r-Rao bounds we have that the Fisher information in Eq. (\ref{FI}) after condition 
(\ref{1c}) and using $P(k= \pm | \phi)$ in Eq. (\ref{ppme}) becomes $F \simeq 4 \bar{n}^2 / \nu^2$ so that the
Cram\'{e}r-Rao bound (\ref{CRB}) is
\begin{equation}
\Delta^2 \tilde{\phi}_{CR}  \simeq \frac{\nu^2}{4 m \bar{n}^2}  .
\end{equation}
This is just equal to the actual Bayesian uncertainty  achieved in this scheme $\Delta^2 \tilde{\phi}$ in Eq. (\ref{ue}) 
so our scheme saturates the Cram\'{e}r-Rao bound. Moreover, it saturates also the quantum Cram\'{e}r-Rao 
bound since the quantum Fisher information in Eq. (\ref{QFI}) for $\nu \ll 1$ matches the Fisher information 
$F_Q \simeq F \simeq 4 \bar{n}^2 / \nu^2$, so that 
\begin{equation}
\Delta \tilde{\phi} \simeq  \Delta \tilde{\phi}_{CR} \simeq \Delta \tilde{\phi}_{QCR} .
\end{equation}

Regarding the Bayesian Cram\'{e}r-Rao bound in Eq. (\ref{BCR}) and considering in our case $\mathcal{I} \simeq
1/W^2$ we get 
\begin{equation}
 m \int d \phi P(\phi)  F + \mathcal{I}  \simeq 4 m \bar{n}^2 / \nu^2 + 1/W^2  \simeq 4 m \bar{n}^2 / \nu^2 ,
 \end{equation}
 where we have used condition (\ref{2c}). Therefore we get that our scheme also saturates this Bayesian 
 Cram\'{e}r-Rao bound, so that 
 \begin{equation}
\Delta \tilde{\phi} \simeq  \Delta \tilde{\phi}_{CR} \simeq \Delta \tilde{\phi}_{QCR} \simeq \Delta \tilde{\phi}_{BCR} .
\end{equation}

\section{Beating the weak Heisenberg limit}

\begin{figure}
\begin{center}
\includegraphics[width=7cm]{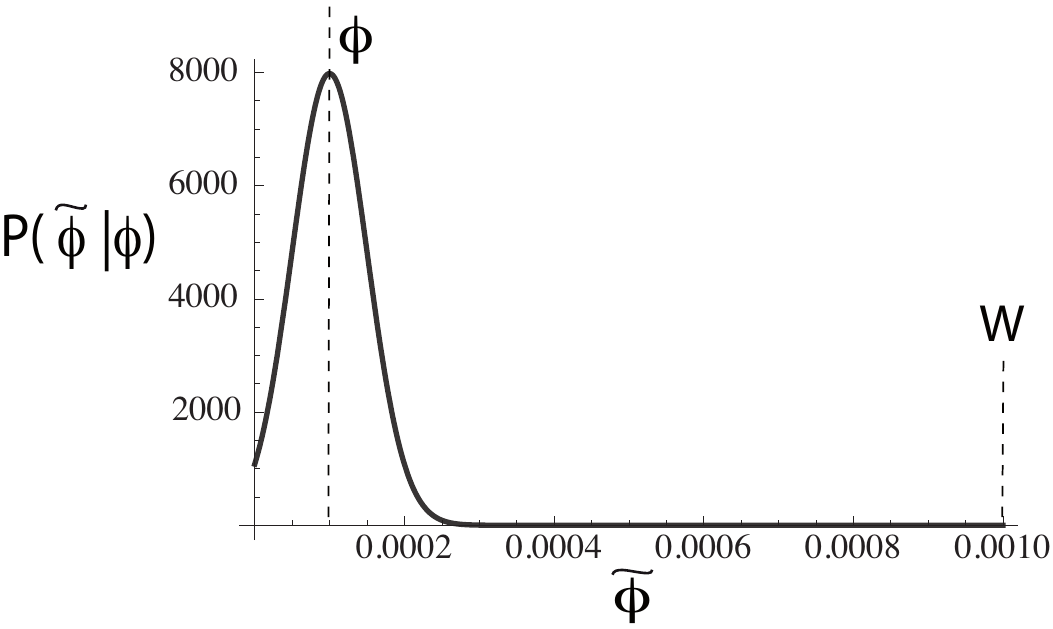}
\end{center}
\caption{Posterior distribution $P ( \tilde{\phi} | \phi  ) $ for the estimator $\tilde{\phi} $
for $W = 10^{-3}$, $\bar{n}=1$, $m= 10^6$, $\nu = 0.1$, and $\phi = 10^{-4}$. The exact and approximate  
expressions are indistinguishable. We have marked with vertical dashed lines the values of $\phi$ and $W$. }
\end{figure} 

Let us show explicitly that the above analysis provides a proof of principle that the weak Heisenberg 
limit can be beaten. To this end we present a numerical evaluation of the posterior distribution without 
any approximation directly by combining Eqs. (\ref{bin}),  (\ref{est}) and  (\ref{post}). This is compared 
with the approximation (\ref{posta}).  In Fig. 1 we have plotted both posterior distributions for  
 \begin{equation}
 W = 10^{-3} , \; \bar{n}=1 , \; m=  10^6, \;  \phi = 10^{-4}, \; \nu = 0.1 .
 \end{equation}
Both conditions (\ref{1c}) and (\ref{2c}) are satisfied with $ W \bar{n} /\nu^2 =0.1$, $\sqrt{m}\bar{n} W /\nu = 10$
and $m \nu^2 = 10^4$. So the exact and approximate expressions are indistinguishable and the uncertainty is 
readily given by Eq. (\ref{ue}). More specifically, in this case we get $\Delta \tilde{\phi} = 5 \times 10^{-5}$ which is 
clearly  below both the weak limit and the prior, being above the strong limit 
\begin{equation}
\Delta \tilde{\phi}= 0.05 \Delta \phi_w = 0.05 W = 50 \Delta \phi_s ,
\end{equation}
where for definiteness we have considered $\Delta \phi_w = 1/(\sqrt{m} \bar{n})$ and $\Delta \phi_s = 
1/(m \bar{n})$. 

Next we can examine whether this strategy can approach the strong Heisenberg limit for a proper choice of parameters. 
This is the case for example of an slight variation of the above example as 
\begin{equation}
 W = 10^{-3} , \; \bar{n}=1 , \; m= 1.6  \times10^4, \;  \phi = 10^{-4}, \; \nu = 0.03,
\end{equation}
leading to  $\Delta \tilde{\phi} = 1.1 \times 10^{-4}$ with 
\begin{equation}
\label{gr}
\Delta \tilde{\phi}= 0.015\Delta \phi_w = 0.11 W = 2 \Delta \phi_s ,
\end{equation}
so this is just twice the strong Heisenberg limit. In Fig. 2 we have plotted both the exact and approximate 
posterior distributions showing again that they almost coincide so Eq. (\ref{ue}) holds. In this case the first 
condition  (\ref{1c}) is not satisfied since $ W \bar{n} /\nu^2 =1.1$ while the second one (\ref{2c}) is close to be 
satisfied as $\sqrt{m}\bar{n} W /\nu = 5$, leading to $m \nu^2 = 20$.

\begin{figure}
\begin{center}
\includegraphics[width=7cm]{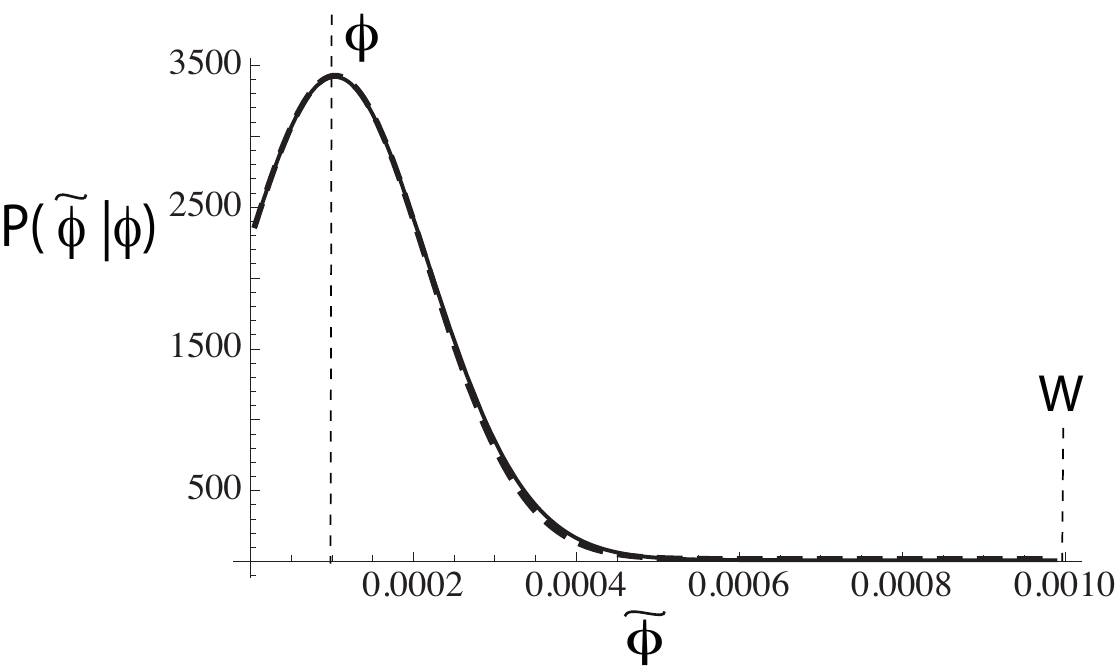}
\end{center}
\caption{Posterior distribution $P ( \tilde{\phi} | \phi  ) $ for the estimator $\tilde{\phi} $
for $W = 10^{-3}$, $\bar{n}=1$, $m= 1.6 \times10^4$, $\nu = 0.03$, and $\phi = 10^{-4}$.
The exact (solid line) and approximate (dashed line) expressions are indistinguishable. 
We have marked with vertical dashed lines the values of $\phi$ and $W$. }
\end{figure} 

 \section{Discussion}

We have provided analytical and numerical evidences showing that the weak form of the Heisenberg 
limit can be beaten while the prior information is improved without bias. The probe and measurement 
presented may be regarded as unpractical, but the main goal was to provide a simple as possible 
proof of principle of the beating. 

The key point is that in any case the weak-probe scenario is better than the bright-probe case. This is 
because special states of light, such as the probe (\ref{probe}) or even N00N states, are much more accessible 
and robust against practical imperfections for small mean numbers that for large numbers. In this regard the use 
of weak coherent states is excluded since they would deprive us of the $\nu$ parameter which has been 
crucial in the above analysis. 

One of the main questions addressed in quantum metrology is how to use resources in the most efficient way. 
The answer is not simple and one must trust the conclusions provided by the performance estimator chosen, in 
this case the mean square error averaged over prior distribution. Coherent states are much easily 
produced but they are not efficient enough according to these statistical tools. We may say that although each 
run provides a minuscule amount of information, it is of a much larger quality regarding long run cumulative effects 
when compared to a single measurements with all photons gathered in a bright coherent state.

Quantum-metrology protocols proceed without explicit references to the physical meaning of the parameter 
to be estimated. Nevertheless, in our case we may question what the phase in Eq. (\ref{ps}) is relative to. This 
point is settled by the measurement process that directly or indirectly embodies the reference phase. We may 
say that this is a somewhat sophisticated version of  homodyne or heterodyne quadrature measurements, 
where the phase of a single-mode field can be suitably observed relative to the phase of the local oscillator 
that defines the actual quadratures being measured.

Saturation of bounds is a quite tricky point in every estimation procedure. Actually, this was the main reason 
for focusing on Bayesian approaches. We have shown that our scheme saturates all bounds 
including the Cram\'{e}r-Rao bounds with quantum and classical Fisher informations. 

Finally, a merit of these results is that they point to a largely dismissed possibility in quantum metrology, this is the multiple 
repetitions in a weak-probe scenario. To secure this point we have shown that it avoids most of the 
difficulties that face more naive approaches such as prior information and bias.  We show that is a quite 
interesting route to be followed to obtain suggestive nontrivial results.
 
\section*{Acknowledgements}
I thank Drs. \'Angel Rivas and Jes\'us Rubio for enlightening discussions
and support from Spanish MINECO grant FIS2012-35583,  and from the CAM research 
consortium QUITEMAD+ grant S2013/ICE-2801.

\end{document}